\documentclass[sigconf]{acmart}
\usepackage{booktabs} % For formal tables

\acmDOI{10.1145/3201064.3201107}

\acmISBN{978-1-4503-5563-6/18/05} 

\acmConference[WebSci'18]{10th ACM Conference on Web Science}{May 27--30, 2018}{Amsterdam, Netherlands}
\acmBooktitle{WebSci '18: 10th ACM Conference on Web Science, May 27--30, 2018, Amsterdam, Netherlands}
\acmYear{2018}
\copyrightyear{2018}
\setcopyright{acmlicensed}
\acmPrice{15.00}

\begin{document}
\title{Pathways to Fragmentation}
\subtitle{User Flows and Web Distribution Infrastructures}
\subtitlenote{Both authors contributed equally to this work.}

\author{Harsh Taneja}
\orcid{0000-0002-4630-8911}
\affiliation{%
  \institution{University of Illinois Urbana-Champaign}
  \city{Urbana}
  \state{IL}
  \country{USA}
}
\email{harsh.taneja@gmail.com}

\author{Angela Xiao Wu}
\affiliation{%
  \institution{New York University}
  \city{New York}
  \state{NY}
  \country{USA}
}
\email{angelaxwu@nyu.edu}

\renewcommand{\shortauthors}{H. Taneja and A. X. Wu}

\begin{abstract}
This study analyzes how web audiences flow across online digital features. We construct a directed network of user flows based on sequential user clickstreams for all popular websites \textit{(n=1761)}, using traffic data obtained from a panel of a million web users in the United States. We analyze these data to identify constellations of websites that are frequently browsed together in temporal sequences, both by similar user groups in different browsing sessions as well as by disparate users. Our analyses thus render visible previously hidden online collectives and generate insight into the varied roles that curatorial infrastructures may play in shaping audience fragmentation on the web.
\end{abstract}




\maketitle

\section{Introduction}

Constellations of websites competing for user attention result in patterns of online audience fragmentation, which has garnered enormous interest. Yet we know little about broader patterns in people's browsing pathways and how these pathways contribute to macro patterns of web use. Hence, in this study, we analyze how users flow across online digital features with a focus on the role of digital infrastructures in shaping online consumption patterns. Studying user flows enriches our understanding of audience behavior in today's digital environment by teasing out various patterns of user trajectories as building blocks of macro fragmentation. It also helps investigate into the roles played by digital platforms with distinct mechanisms of traffic curation, such as those of social media, search engines, and web portals \cite{napoli14}.

\section{USER FRAGMENTATION}
Although there's general agreement that web audiences are fragmented in their patterns of exposure, opinions are divided on its underlying mechanisms and consequences. Given that people have a lot of choice, one school of thought believes that people are now free to exercise their preferences and media actors have less control over what people get exposed to \cite{prior07,stroud10}. In this vein, we would expect web users to retract into their own echo chambers, or their online  ``filter bubbles'' \cite{pariser11}. Studies at the user level, which focus on people's engagements with specific online content and digital platforms, typically advance such findings. 

Others argue people's online behavior often does not reflect their self-reported preferences \cite{webster14}. This is because, most people, despite their divergent preferences tend to gravitate towards a handful of popular items due to media producers promoting these aggressively as well as under social influence. This view emanates from studies that analyze macro patterns of web use and are based on snapshots of shared consumption between outlets \cite{wu16}. The focal unit in latter studies is audience duplication which, simply put, is the extent to which two media outlets (e.g., websites) are consumed by the same set of people in a given time period.  In a hypothetical universe of 100 people, if on a given day 20 people accessed both CNN.com and Google.com, the audience duplication between these two outlets would be 20 or 20\%. Examining pairwise duplication simultaneously for all possible pairs of media outlet, these provide insights into outlets commonly consumed together in the aggregate.  For instance a study found that in the US in 2009 most outlets across the 200 most used TV channels and websites had high audience overlaps, contrarian to the popular worry about online polarization \cite{webster12}. 

We argue that in addition to the efforts by media actors and user preferences, technological infrastructures are foundational to shaping online user behavior. On the web, both media actors and users simultaneously utilize multifarious ensembles of curatorial practices. For instance, a user active on online social networks may access news articles by a particular publisher who more regularly promote their news through its pages on social networks. Another user who has a portal as a homepage may tend to access news from publishers that make their news available on that portal. In other words, there are just myriad such curated flows that influence people's exposure to content \cite{thorson16}. To unravel these requires a novel empirical strategy, which we develop in this study. 

\section{ANALYZING CLICKSTREAMS}
Analysis of audience duplication does provide a holistic snapshot of media consumption patterns in a high choice environment. However, it is a static snapshot unable to capture the dynamics of audience flows leading to such patterns. We elaborate on this with an example. Consider two partisan news outlets Fox News and CNN. Audience duplication data have consistently shown greater than expected audience overlaps between their TV channels \cite{webster05} as well as their websites \cite{gentzkow11}. However, this high duplication between these two ideologically opposing outlets may result from different pathways.  Over the time period of analysis, people may be directly visiting these outlets (in succession) to ``check out the other side,'' or they may be directed to both these outlets from search engines or social networks. But discerning these differences in pathways is beyond the radar of audience duplication analysis. 

To overcome this limitation, we demonstrate a ``sequential'' audience centric approach to audience fragmentation that uses clickstream data to unravel audience flows. Clickstreams are traces of user activity as they move from browsing one webpage to the other \cite{wu14}. Analyzing these clickstreams alongside audience duplication, our sequential approach unpacks the aggregated shared traffic between websites by teasing out the constituent browsing sequences users undertake across the web. This approach has two major merits: (1) By examining groups of websites that are associated with one another in terms of recurring traffic flows, we render visible previously hidden online populations. These collectives are distinct not (merely) by demographic profiles or commonly used online outlets, but by sequential movements on the Web. (2) Linking the patterns of sequential browsing sessions we observe to existing debates about the "curatorial mechanisms" of sites such as search engines and social media, we are able to further discern the nature of these sequential sessions. This generates insight into the varied roles that curatorial infrastructures may play in shaping audience fragmentation online. 

\begin{figure}
\includegraphics{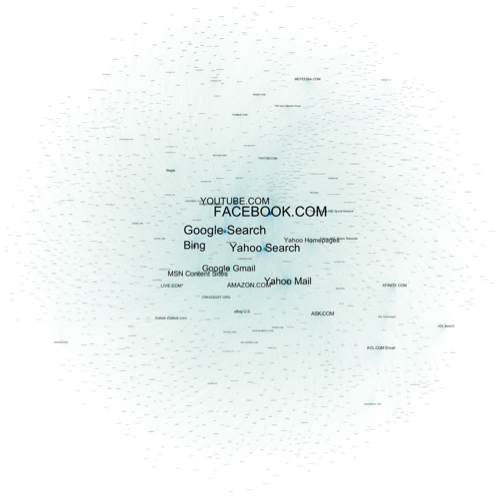}
\caption{Clickstream network (1761 nodes, 89453 edges)}
\label{fig:1}
\end{figure}

\subsection{Data}
In clickstream data, when a user on outlet i switches to outlet j and then to outlet k, outlet i is a source of traffic for j, which in turn is a source of traffic for k. Aggregating all these clickstreams in a media ecosystem, one can deduce how audiences ``flow'' at a macro-scale. 

We obtained US national level web usage (traffic) data from comScore. Our sample includes all the 1761 web outlets visited by at least 1 \% (2.6 million) of all US web users in October 2015. For each web outlet, we extracted its clickstream data, which reflects the number of total unique users during the said month that landed on it immediately after visiting other outlets. Owing to the large and expansive sample size, for most sites, we had the complete extent of their sources of incoming traffic. We aggregated these data for all websites to create a User Flow Matrix, which is a directed network with web outlets as nodes and the user volumes from one outlet to another as the edges. We also obtained the pairwise audience duplication between all 1761 outlets, which for any outlet pair is the percentage of users that visited both. 

\subsection{Analysis and Results}

\subsubsection{Clickstream Volumes}
First, we calculated the weighted in-degrees and out-degrees of each node. In this network, a node's weighted out-degree represents the total volume of user traffic that flows out of this site to all other sites in the network. Although both distributions are quite skewed leading to a highly centralized network, the latter are highly concentrated (Gini Coefficient = 0.64). Thus, in the aggregate, we find that most web traffic flows outwards from an extremely small number of web outlets. We visualize this graph using force atlas algorithm, which also takes into account tie weights (see Figure \ref{fig:1}). The nodes are sized according to their out degree weights and we can see the largest sized nodes which are also at the center of the graph are the most popular search engines, email providers, social networks and portals. Their large out-degrees suggest that these handful sites precede whichever other sites users visit on the web. 

\subsubsection{Clickstream Constellations}

\begin{figure}
\includegraphics{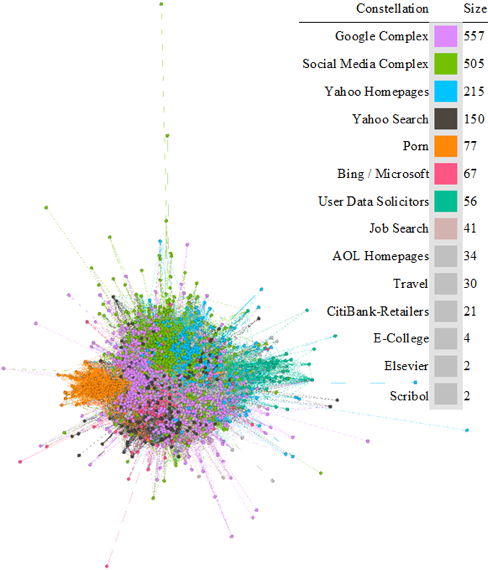}
\caption{14 Clickstream constellations}
\label{fig:2}
\end{figure}

We conducted a cluster analysis using a modularity based community detection algorithm \cite{blondel08}. Modularity based clustering isolates groups of websites with dense interconnections within groups relative to edges between groups. A cluster in this network contains pairs of websites that users tend to access in succession, which may be seen as an aggregation of socially shared browsing sequence, which we refer to as ``constellations.'' Consistent with our research objective, we accepted a cluster solution where we could identify for each constellation either content similarity or some underlying digital infrastructures (including customized curation embodied by search engines and social network sites, as well as top-down embedded architectures such as web portals). Our solution, shown in Figure \ref{fig:2}, revealed 14 clusters or constellations, which represent distinct patterns of web browsing trajectories. Owing to the small number of websites and highly specialized nature of their content we did not consider the three smallest clusters for further analysis. 

To determine the extent to which a handful of sites ``anchored'' these browsing sequences, by serving as starting and returning points during browsing sessions, for each cluster we calculated a Gini coefficient (reported in Table \ref{tab:freq}) based on the distribution of the weighted out-degrees of its constituent sites. A high Gini coefficient indicates that a handful of sites anchor the browsing sequence among constituent sites--that is, within the browsing sequence, whichever websites the users land on, their preceding visits tend to be on these anchor sites. Given how people usually browse the web, it is more likely that they move from anchor sites to the subsequent site through clicking embedded links instead of deliberately entering the subsequent site's URL. To further discern the likely mechanisms, we examine the nature of the anchors (those with the highest weighted outdegrees in each constellation, also listed in Table \ref{tab:freq}). It turns out that most of these anchors embody certain infrastructural designs. As discussed, anchors in high-Gini constellations are more influential in shaping the browsing sequences within each constellation. 

\begin{table}
\small
  \caption{Analyses of clickstream distributions}
  \label{tab:freq}
  \setlength{\tabcolsep}{0pt}
        \begin{tabular}{ccc}
    \toprule
    Constellation&Gini (Out-degrees)&Anchor(s)\\
    \midrule
    Bing/Microsoft & 0.89& Bing, MSN Content Sites\\
    Google Complex & 0.85& Google Search, Youtube, GMail\\
    Social Media Complex & 0.83 & Facebook\\
    Yahoo Homepages & 0.82 & Yahoo Homepages, Yahoo Mail\\
    Yahoo Search & 0.81 & Yahoo Search\\
    AOL Homepages & 0.72 & AOL Homepages, AOL Email\\
    User Data Solicitors & 0.62 & Swagbucks\\
    Citibank-Retailers & 0.62 & Citibank, Macy\\
    Porn & 0.61 & Pornhub\\
    Job Search & 0.56 & Indeed\\
    Travel & 0.43 & Tripadvisor, Expedia, Priceline\\
  \bottomrule
\end{tabular}
\end{table}

\textit{Google Complex}: With the largest number of sites, this cluster mainly contains utilitarian websites such as those of retailers (including Amazon and eBay) along with those of service providers in domains such as government, finance, travel, telecom and shipping. One common feature about all these sites is that users visit them to accomplish a particular purpose and are unlikely to encounter them as part of random browsing. Its highest outdegree suggests that users rely on Google Search to visit most constituent websites. Notably, YouTube and Gmail are also main anchors, suggesting they receive intermittent visits during typical browsing sessions. 

\textit{Social Network Complex}: This cluster has the most popular online social networks such as Facebook (also the anchor with impregnable dominance), Instagram, Twitter, and LinkedIn, among others. Also included are a bulk of symbolic content spanning news, sport and entertainment sites both by legacy media organizations (e.g., CNN and New York Times) as well as digital native providers (e.g., Buzzfeed and DrugeReport). Further, there are ``socially-driven'' outlets such as GoFundMe, Spotify, Fitbit, and Legacy.com, and sites of various banks and mobile/ISPs that long-term customers most likely access directly amidst their social media browsing through bookmarks or typing the URL, without turning to search engines. 

\textit{Yahoo Homepages}: A third somewhat smaller cluster contains a variety of Yahoo's biggest online properties (e.g., Yahoo Homepage, Yahoo News, Yahoo Sports, and Yahoo Mail). It also has many sites specialized in elaborate political news commentaries (e.g., Vox.com, Slate, Atlantic, Politico, Dailybeast, TheHill), general ``soft'' news (e.g., Huffpost and USAToday), business news (Bloomberg and Business Insider), online services for investments and mortgages, as well as Classmates.com that helps search for past high school friends.  

\textit{Yahoo Search}: Other than Yahoo Search, this cluster has many other utilitarian websites most of which serve a specific purpose and originated in the 1990s including as Ask.com, Ehow, MapQuest and WedMD. This cluster probably represents a segment of web users that still use Yahoo Search as their gateway to the web and are presumably older than the average web user. 

\textit{Porn Constellation}: This is a cluster comprised of adult sites. Owing to its high outdegree, our analysis suggests that for most users Pornhub.com serves as a gateway to other adult websites, many of which (e.g., Youporn and Xtube) are linked to by Pornhub.com's homepage and are part of the ``Pornhub Network.'' Many of the other adult sites that feature in this cluster appear to regularly advertise their content through adult advertising networks on the Pornhub network sites. 

\textit{Bing / Microsoft}: We also observe a cluster (with the highest Gini) anchored by Microsoft's search engine Bing. It consists of many other web services and content portals owned or aligned with Microsoft, such as Office, Windows or AccuWeather (Bing's default weather widget). 

\textit{User Data Solicitors}: This cluster contains content that solicits user inputs for commercial data extraction through ads or popups on other websites. Constituent sites include online survey interfaces, rewards and discount solicitations as well as some of the market research companies' websites that design these surveys and analyze the collected user responses. 

\textit{Job Search}: A group of popular job search and application portals such as Indeed.com and Monster.com, as well as websites of companies that job portals use for database management. 

\textit{AOL Homepages}: This is a relatively small cluster anchored by ``AOL Homepages,'' which contains a few AOL websites (e.g., AOL Email). It also has sites such as USMagazine, EverydayHealth, and ZergNet, all part of the AOL brand. 

\textit{Travel}: With TripAdvisor as the biggest anchor, this cluster contains sites specialized in booking flights, hotels, and travel packages along with hotel and airline websites. 

\textit{Citibank-Retailers}: This cluster consists of Citibank and many online retailers such as Macy, K-Mart, OldNavy, and Sears, all of which have credit card programs managed by Citibank.

\subsubsection{Audience Duplication between Constellations}
\begin{figure}
\includegraphics{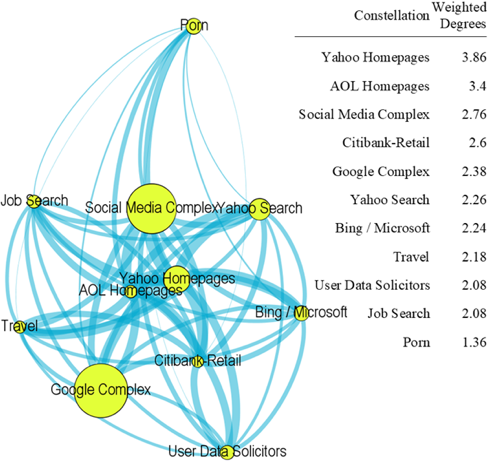}
\caption{Audience duplication between constellations}
\label{fig:3}
\end{figure}

To discern the degree to which each clickstream constellation shares users with other constellations, we computed the greater than expected (random) audience duplication between each website pair. Then we averaged these pairwise duplication figures for each constellation pair. For example, for two constellations with a site count of ``m'' and ``n'' respectively, we averaged the greater than expected duplication for each of the constituent ``m*n'' website pairs. This resulted in an 11*11 symmetric matrix, where each cell indicates the average greater than expected duplication between websites at the level of constellation pairs. Figure \ref{fig:3} plots this matrix as a force directed network map using the Force Atlas 2 layout in Gephi, which also considers the edge weight.

In Figure \ref{fig:3}, each node is a constellation, sized according to its number of websites. The weighted degrees on each node indicated the extent to which a constellation's constituent websites share users with websites in other constellations.  The edges with the highest edge weights in this network are those between the three constellations with highest weighted degrees. The top edges are, respectively, those between Yahoo Homepages and AOL Homepages, Social Media Complex and Yahoo Homepages, and AOL Homepages and Social Media Complex.

\section{A SEQUENTIAL EXPLANATION }

\subsection{In-constellation Fragmentation}
Within each constellation, we examined the curatorial characteristics of major anchoring sites in relation to other constituent websites in light of the constellation's general distribution of traffic flows. This revealed how the different mechanisms of the anchors' digital curation shape online user fragmentation and online browsing routines at large. 

First, our analysis suggests that social media and search engines, which embody two distinct mechanisms in curating traffic, impact different parts of the web. A social media site directs users as it runs its curatorial algorithms on content updates generated by user's personal network. A search engine, in contrast, directly responds to requests submitted by individual users. We found that websites in search engine-anchored clickstream constellations tend to differ in nature from those anchored by social networks. The former are largely websites for retail, banking, everyday services such as map and health consultation. In contrast, the latter tend to be websites providing news, commentaries, and entertainment content, basically information as an end in itself. This differentiation is one between utilitarian, functional goods and symbolic, experiential goods. Our results thus suggest that people tend to land on news information sites largely through social networks (the Facebook Constellation include multiple other popular social media sites). Driven by social media use, the nature of such encounter is most likely incidental. In contrast, people tend to fulfill their everyday tasks by deliberately seeking services via search engines. 

Second, the composition of clickstream constellations anchored by web portals suggests strong evidence of infrastructural bundling, by which we mean conglomerates using architectural design to foster people already on one of their sites to visit other sites owned by the same companies. For example, both the AOL and the Yahoo Constellations include not only bundles of the whole spectrum of the brand's online services and content offerings, but also discrete websites with underlying capital connections to the brands. The Porn Constellation is another striking example, where by design niche adult sites owned by the MindGeek conglomerate form a closed maze that confines sequential users' flows. 

Less visible to users, a third type of traffic curation is more of a function of back-end infrastructural bundling due to inter-organizational arrangements. For example, Citibank-Retailer comes into being because all the retailer sites have Citibank as their payment gateway. Likewise, the Job Search Constellation is stitched together due to the various job portals managing their data through common vendors such as Salesforce or Oracle. 
In summary, while search engine usage entails some user initiation, other anchoring sites ranging from social media, web portals, and contracted technical platforms on the backend seem to direct traffic in ways that require decreasing levels of conscious planning and participation on the part of the user. 

\subsection{Across-constellation Fragmentation}
We used audience duplication between website pairs in different constellations to discern the extent to which they are used by similar or dissimilar user groups. Viewed in light of the nature of anchors, this result illuminates broad contours of fragmentation across the user bases of distinct constellations. 

To begin, we indeed observed evidence of content preferences of what seem specific user niches. The relatively isolated constellations, Porn, Job Search, and User Data Solicitors point to particular populations by gender, age, and class. The Solicitors Constellation, for example, with its notable connections to Citibank-Retailers, and Yahoo and AOL Homepages, suggests browsing activities of a middle-aged homemaker user group. That said, bulk of the reported evidence shows that various infrastructural factors, on which we elaborate next, can explain across-constellation fragmentation. 

First, routines of everyday web use, usually neglected in extant research, come to the fore. Specifically, ``anchor'' websites are incorporated into people's browsing routines  based on the sites' distinct functionalities along the infrastructural dimension. "Infrastructural functionality" explains the lower mutual user overlaps between sites in the three constellations driven by search engines (i.e., Google Search, Yahoo Search, and Bing), as users typically rely on one search engine, the algorithm of which comes to be increasingly responsive to their individual behavior. This result further suggests that people tend to fragment into using different clusters of websites respectively anchored by different engines. Infrastructural functionality also explains the fact that, relative to sites in search engines-anchored constellations, sites in the Social Media Complex has higher tendency to share users with sites in constellations anchored by general interest portals such as Yahoo Homepages and AOL Homepages. This is because for typical web users, it is habitual to linger on and around social network sites and share content to which portals also link. 

We also observe effects of infrastructural bundling across constellations, further highlighting the striking influences of commercial packaging by design. Notably, sites in the AOL Homepages and Yahoo Homepages constellations have the highest user overlaps, most likely driven by their content partnerships. The sports section on AOL is populated by ``Yahoo Sports'' and HuffPost (an AOL company) provides business and lifestyle news on the Yahoo.com home page. More recently, Yahoo and AOL have both been incorporated into one company, Oath Media. 

Finally, we found significant portions of users browsing sites in the Bing/Microsoft Constellation also browse those in AOL and Yahoo Homepages. This suggests the influence of ``infrastructural residuals,'' or effects of enduring infrastructural features in shaping media habits \cite{taneja17}. This simultaneous engagement with sites across these constellations points to the collective habitat of a senior user population that has been habituated into the reliance on AOL and Yahoo sites (both long-standing web brands); a substantial portion of this population also uses Bing due to technical prompts as Bing is set as the default landing website for the age-old IE browsers. Infrastructural residuals thus lead to fragmentation by generation as a result of temporally variant habituation with new media.

\section{CONCLUSION}

Our results show that there seems to exist one relatively expansive user population whose online life is anchored by Google Search, one (presumably relatively older) population who relies on portals such as Yahoo and AOL homepages and a third user group that defaults to Microsoft and uses Bing. Importantly, this fragmentation results from an interaction between users intentions and structural conditions including infrastructural functionality, infrastructural bundling, and infrastructural residuals. After users embark on major anchoring sites, different curatorial architectures that foster specific browsing sequences further user fragmentation. 

In sum, viewed in the aggregate, rather than subjective preferences, the infrastructural dimension explains a great extent of online fragmentation. In fact, it owes much of its power to its invisibility to the user. Furthermore, infrastructural factors that have arisen from the established political economy configuring the internet industry and media industries at large appear to shape online fragmentation more resolutely than the much-debated ``customizing technologies'' such as social networks and search engines.

%\end{document}  % This is where a 'short' article might terminate

\bibliographystyle{ACM-Reference-Format}
\bibliography{sample-bibliography}

\end{document}